\documentclass[journal]{JPercepImag}

\usepackage{graphicx,color,rotating,calc,xcolor}
\usepackage{cite}

\ifCLASSOPTIONcompsoc
  \usepackage[caption=false,font=normalsize,labelfont=sf,textfont=sf]{subfig}
\else
  \usepackage[caption=false,font=footnotesize]{subfig}
\fi

\usepackage{moreverb,url}
\usepackage[colorlinks,bookmarksopen,bookmarksnumbered,citecolor=red,urlcolor=red]{hyperref}

\newcommand\BibTeX{{\rmfamily B\kern-.05em \textsc{i\kern-.025em b}\kern-.08em
T\kern-.1667em\lower.7ex\hbox{E}\kern-.125emX}}

\usepackage{listings}
\usepackage{color}
\definecolor{lightgray}{rgb}{.9,.9,.9}
\definecolor{darkgray}{rgb}{.4,.4,.4}
\definecolor{purple}{rgb}{0.65, 0.12, 0.82}
\lstdefinelanguage{JavaScript}{
  keywords={break, case, catch, continue, debugger, default, delete, do, else, false, finally, for, function, if, in, instanceof, new, null, return, switch, this, throw, true, try, typeof, var, void, while, with},
  morecomment=[l]{//},
  morecomment=[s]{/*}{*/},
  morestring=[b]',
  morestring=[b]",
  ndkeywords={class, export, boolean, throw, implements, import, this},
  keywordstyle=\color{blue}\bfseries,
  ndkeywordstyle=\color{darkgray}\bfseries,
  identifierstyle=\color{black},
  commentstyle=\color{purple}\ttfamily,
  stringstyle=\color{red}\ttfamily,
  sensitive=true
}

\newcommand{\putfigman}[2]{\includegraphics[width=#2 \columnwidth]{#1}}

\begin{document}

\title{Sketch-and-test: picture-centered research with p5.js assisted crowdsourcing.}
\author{M.W.A. Wijntjes and M.J.P. van Zuijlen
\thanks{M.W.A. Wijntjes and M.J.P. van Zuijlen both work at the Perceptual Intelligence Lab, Delft University of Technology, Landbergstraat 15, 2628CE Delft, The Netherlands.}
 \thanks{Corresponding author: m.w.a.wijntjes@tudelft.nl}
 \thanks{Repository: https://github.com/maartenwijntjes/p5-sketch-and-test}
}
 \maketitle

\section{Abstract}

Relating human judgements to pictures is central to a wide variety of scientific disciplines. Pictures are used to evoke and study faculties of the human mind, while human input is used to label, understand and model pictorial representations. 

Human input is often collected through online crowdsourcing experiments. This paper discusses the usage of crowdsourcing in two major branches of picture-centered research, human and computer vision, and identifies novel directions such as art history and design. We demonstrate that a wide variety of experiments can be conducted by using p5.js, a library originally intended to facilitate visual creation.

We report five complementary experimental paradigms to illustrated the accessibility and versatility of p5.js: Change blindness, BubbleView, 3D shape perception, Composition, and Perspective reconstruction.  Results reveal that literature findings can be reproduced and novel insights can easily be achieved with the p5.js library. 

The creative freedom of p5.js combined with low threshold access to crowdsourcing  seems like a powerful combination for all picture-centred research areas: perception, design, art history, communication, and beyond. 

\section{Introduction}

The visual analysis of pictures plays an important role in a large variety of scientific areas, such as art history, sociology, anthropology, psychology, arts and design. Often, this research is `qualitative' (e.g., \cite{margolis2011sage,rose2016visual}), as quantitative approaches either do not exist or have a too narrow scope for the research questions. When taking art history as an example, the qualitative approach concerning the representation (e.g., iconography or formal analysis) contrasts strongly with the amount of quantitative analysis that is spent on the physical object. Pigments are mapped, canvas weavings identified and under drawings scanned, while the \emph{representation} is left for qualitative research. 

Not \emph{all} research that involves pictures is qualitative, but scientific fields using a quantitative approach also often have different research questions. Experimental psychology uses pictures extensively, but not so much for the pictures' sake, but rather for the beholders' response. Pictures are used as `stimuli' that evoke perceptions and controlling certain aspects allows for inferences about these perceptions. The experimental paradigms used are as extensive as the perceptual mechanisms under investigation: attention, colour, depth, perceptual organisation, texture perception etc. 

Another field that has a quantitative approach towards images is computer vision. To create computer vision models, training and test data is needed that consists of both the image and some form of metadata such as object category, but also more complex annotations. Both computer and human vision practices rely on human input: human vision is understood via behavioural experiments while computer vision uses image annotations as ground truth data. Both have been using online crowd sourcing platforms extensively over the past few decades. Both behavioural and annotation experiments require a certain level of accuracy and versatility. A similar desire is shared by visual artists and designers who developed a Processing inspired JavaScript library called p5.js. In this paper we aim to demonstrate how p5.js can be employed for perception and annotation experiments, and also demonstrate potential extensions towards other fields such as art history. We believe that picture-centered research shares similar experimental paradigms while having different research questions. By focusing and these paradigms and explaining how they can relatively easily be conducted online, we aim to be of practical use but also contribute to a conceptual level with respect to picture-centered research. 

Before discussing and demonstrating p5.js, we will first review visual crowdsourcing research for human and computer vision.

\subsection{Perception research}

Collecting data through online crowd experiments has become a standard research practice for the behavioural sciences. There are various possible motivations to choose online over lab experiments, such as speed and efficiency, but also more specific reasons such as access to certain subject pools \cite{paol2010,buhrmester2016amazon}. Performance seems generally not to be degraded \cite{Buhrmester2011} although participants sometimes lack attention \cite{Goodman2013}. Most paradigms from experimental psychology give similar results when conducted online as compared with lab experiments \cite{Crump2013,Haghiri2019}.

Crowdsourced experiments range from simple questionnaires to complex visual presentations. Within the wide spectrum of behavioural sciences that use crowdsourced experiments, visual perception researchers  are particularly concerned about controlling the presentation, e.g., colour, size, and timing. Relating physical characteristics of the stimulus to subjective experience is an essential element of psychophysics. To this end,  software used by vision scientists is often very customizable and allows for high levels of control. Two packages regularly used are Psychtoolbox \cite{Kleiner} (PTB, formerly called Psychophysics Toolbox \cite{brai1997})  for Matlab  and PsychoPy \cite{Peirce2019} for Python. PTB's own description reflects that vision scientists require both accuracy and versatility to generate stimuli and collect behavioural data:
\begin{quote}
\emph{The PTB core routines provide access to the display frame buffer and color lookup table, reliably synchronize with the vertical screen retrace, support sub-millisecond timing, expose raw OpenGL commands, support video playback and capture as well as low-latency audio, and facilitate the collection of observer responses.}
\end{quote}

While PTB is only available offline, PsychoPy can be run online. To this end, a platform is needed that hosts the JavaScript files generated in PsychoPy and, more importantly, saves the collected data. PsychoPy users can use the platform Pavlovia.org where it is possibly to upload HTML/JavaScript code. Conducting psychophysical studies using JavaScript does not seem to affect control over reaction times \cite{DeLeeuw2016}, a typical dependent variable for vision research\footnote{The library used for this research was jsPsych \cite{de2015jspsych} and may not generalize to p5.}. The third step, after coding and hosting an experiment, is recruiting participants, which can be done through crowdsourcing  marketplaces, such as Amazon Mechanical Turk (AMT) but also many other alternatives.

These three steps (interface/experiment design, online hosting  and participant recruitment) are inherent components of any online experiment. Instead of using a dedicated toolbox, it is possible to directly code an HTML/JavaScript experiment and host it on a web server. Yet, all steps seem relatively high threshold and to this end, a number of initiatives have been taken to facilitate the behavioural scientist. For example, psiTurk \cite{Gureckis2016} is an open source platform that allows researchers to code and use full experiments. One of the motivations behind psiTurk is reproducability: by publishing an online experiment, it can easily be reproduced by peer scientists. One difference between psiTurk and PsychoPy is that psiTurk integrates all three steps of coding, hosting, and recruitment. Another initiative is TurkPrime \cite{Litman2017} which focuses more on  participant management and AMT interface for the researcher.

\subsection{Annotation research}
Online annotation studies share many similarities with perception experiments, except for the rationale: an annotation study is primarily interested in the image and uses the human as an (intelligent) measurement device. Most visual annotation studies originate from computer vision and are aimed at creating databases for machine learning. A well-known example is LabelMe \cite{russell2008labelme}, which aimed at object labeling and polygonal segmentation (where the object is manually outlined). Today, these paradigms are default options available on AMT, but at the time LabelMe facilitated many other segmentation and annotation studies. Besides plain segmentation,  a considerable amount of `richly annotated database' studies have been conducted. These studies often involve the labeling/manipulation of 3D data such as position \cite{dai2017scannet}, 3D meaning of sketch lines \cite{gryaditskaya2019opensketch}, 3D surface attitude  \cite{gingold2012micro,Chen_2020_CVPR} but also material reflectance \cite{bell2013opensurfaces} and shadows \cite{kovacs17shading}. In short, these annotations are often used to retrieve information from pictorial space (i.e., the representation), or to link pictorial space to picture primitives such as lines. From an interface point of view, these experiments share many similarities with perception studies. For example, \cite{Chen_2020_CVPR} annotated surface normals which bears much resemblance to the attitude probe used by \cite{koen1992}. Another example is the material probe used by \cite{bell2013opensurfaces}, who let observers manipulate gloss parameters well known in perception literature \cite{pellacini2000toward}. 

Although computer scientists may have no need for an entry-level library like p5.js that helps them program their annotation tools, it is foreseeable that in the near future annotations will not be reserved to computer science. Firstly, the accessibility of training neural networks will undoubtedly increase, but being able to generate richly annotated training data in a simple fashion is something unexplored. Secondly, annotations can also be used to analyse image collections, for example in the area of digital art history. Here, much activity is focused on the computerized analyses of art collections, but human annotations \cite{Wijntjes2018AnnotatingSH} can also help extract artistic visual information.

\subsection{Visual crowdsourcing with p5.js}

Despite the different motivations underlying perception and annotation studies, they share a similar strategy for data collection. Typical requirements for visual crowdsourcing research are shared by visual artists and designers: they aim to create rich visual experiences with complex user interactions. A well-known programming language used by visual artists is Processing. It aims to be accessible to non-programmers and contains a wide variety of graphical possibilities and options for user interaction. A JavaScript library having much of the functionality of Processing, p5.js, launched its first official (i.e., 1.0) version 1 year ago (February 2020). Over the past few years, beta versions of p5.js \cite{McCarthyLauren2015} have attracted a large community of creative coders ranging from beginners to professional artists. 

\begin{figure}[!h]
\begin{center}
\putfigman{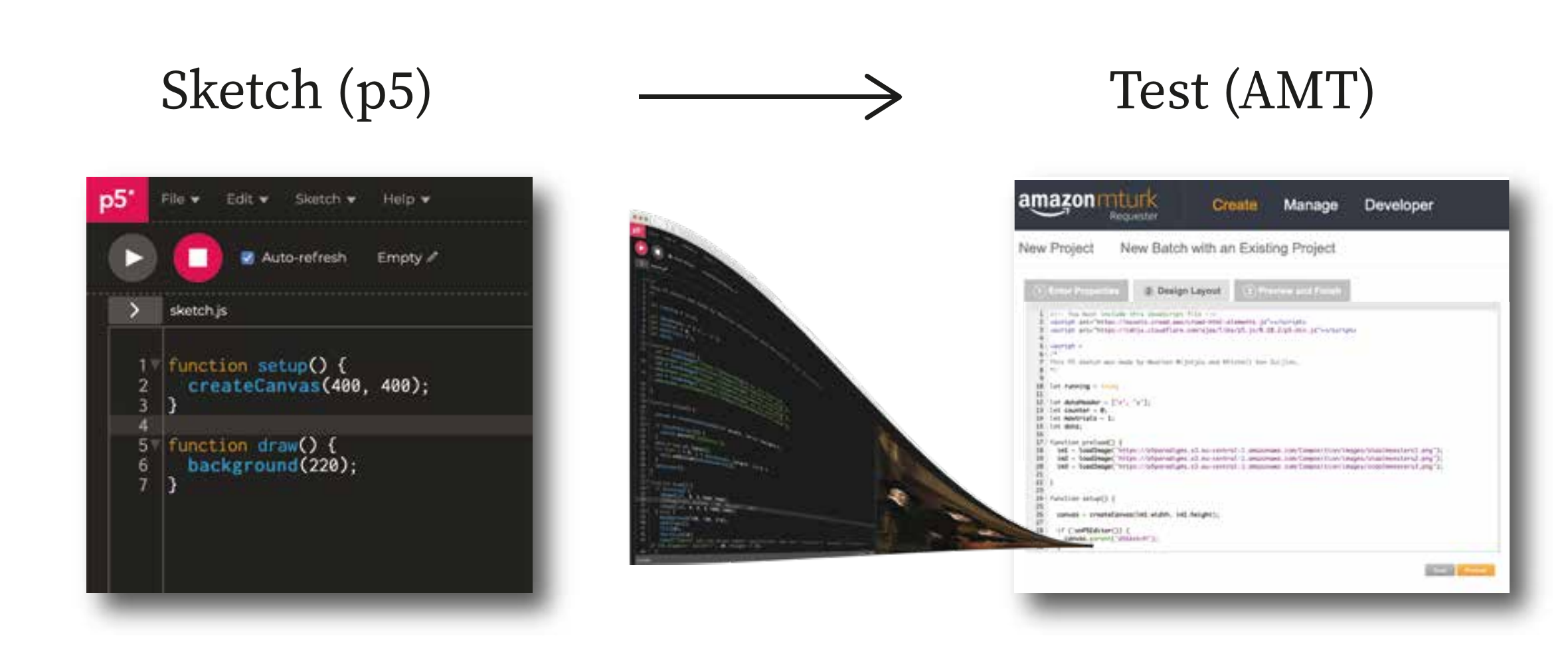}{1}
\caption{The envisioned flow of doing visual crowdsourcing research: first sketch the perception/annotation experiment in the online editor of p5.js, and then copy that code directly into the AMT console.}
\label{visualcrowdresearch}
\end{center}
\end{figure}

Developing stimuli and experiments with p5.js is particularly accessible through the online editor (https://editor.p5js.org). With only some small additions code from the editor can be copy-pasted into the AMT requester portal. This simple flow of quickly `sketching' the experiment in p5.js and than `testing' on AMT (figure \ref{visualcrowdresearch}) can be beneficial to the growing visual crowdsourcing community. We will exemplify the creative freedom and ease of creating an AMT task (HIT, Human Intelligence Task) by reporting five short studies. It should be noted that these experiments demonstrate the use of p5.js, but not \emph{validate} facets such as timing accuracy and usability. For timing accuracy, we believe that p5.js may first show itself useful in time uncritical situations before it can be integrated in benchmarks such as \cite{Bridges2020}. Furthermore,  usability testing is beyond the scope of this paper. Rather, we believe the contribution of this paper lies in a demonstration of doing picture-centered research with p5.js  using classic and novel experimental paradigms.

\subsection{Experiments overview}
We used five different experiments that reflect a variety of interface/interaction designs which are also related to varying research questions. The first experiment is concerned with \emph{change blindness} and makes use of the flicker paradigm \cite{Rensink1996}. While the effect is striking and theoretical implications notable \cite{o1992solving}, a change blindness experiment is very easy to create. In the original study, two pictures are shown subsequently for 240 ms with a gray frame presented intermittently for 80 ms  \cite{Rensink1996}. Besides of fundamental interest, the flicker paradigm also applies to interface design \cite{Varakin2004} and to optimize graphics algorithms \cite{Cater2003}.

A similar deployability for both fundamental and applied research questions can be found in \emph{BubbleView} \cite{Kim2017a}, our second experiment. BubbelView was designed as an online alternative for eye-tracking studies. The paradigm is simple: let observers try to understand what is represented in an image by sharpening selected areas ("bubbles") of a blurred image. Image `understanding' is tested by letting participants describe the picture verbally, while the visual information is quantified by their mouse clicks to sharpen areas. The rationale behind this approach is that it functionally mimics gaze behaviour: where you plan to look (a fixation) is based on low resolution information from your periphery (i.e., outside the fovea).

The third experiment concerns 3D shape perception \cite{Todd:2004lr}. An important experiment in this area  became the so-called \emph{`gauge figure task'} \cite{koen1992}.  It lead to fundamental insights about mathematical transformations describing perceptual ambiguities \cite{Koe2001} and led to more applied studies on the relation between line drawing technique and 3D shape perception \cite{cole}. This method samples the perceived normal vectors of surfaces and is also used as an annotation \cite{gingold2012micro,chen2017surface,Chen_2020_CVPR}.

In the fourth experiment we leave the traditional perception experiments and employ p5.js to answer a question of \emph{artistic composition}. The \emph{Syndics of the Drapers Guild} by Rembrandt is an intriguing painting: the unconventional viewpoint; the feeling of interrupting a meeting that just started. Moreover,  x-ray studies have shown \cite{VanSchendel1956} that Rembrandt had doubts on where to position the servant, the person in the middle behind the others. According to \cite{VanSchendel1956}, Rembrandt initially planned to position this person at the far right of the painting. Informal conversations of art historians led to an interesting question: where would a naive viewer position the servant, and how does this compare to Rembrandt's choice and rejection? This is a typical experiment that can be quickly designed in p5.js and copied into the AMT requester portal.

The last experiment is also related to art history but now concerns more `objective' data in comparison to the aesthetic preferences of experiment 4. We reconstructed \emph{perspective elevation} using simple annotations of the horizon and human figures. Based on the principle of the `horizon ratio' \cite{rogers1996horizon} it is possible to reconstruct the elevation of the viewpoint \cite{Wijntjes2018AnnotatingSH}. We choose to analyse the paintings of Hendrick Avercamp, a Dutch painter from the early 17th century. Avercamp was deaf and dumb (not able to communicate via speech). These disabilities apparently did not obstruct his rise to fame. He is particularly famous for his winter landscapes, often depicting people ice-skating, which happen to be particularly suited for perspective reconstruction because the ice is a horizontal plane.

\section{Methods}
\subsection{p5.js \& AMT}
The name p5 originates from Proce55ing, an alternative spelling of Processing, which is a widely used programming language \cite{reas2007} ``in the context of visual arts''\footnote{https://processing.org}. P5.js is a JavaScript library that  shares much of the functionality of Processing \cite{McCarthyLauren2015}. Both Processing and p5.js aim to make code accessible to a wide audience and are thus relatively easy to use by beginners. A program made by a user is called a `sketch', emphasizing the iterative design process with immediate visual feedback. For a general introduction about p5.js we recommend visiting their webpage\footnote{https://p5.js.org}. Here, we will discuss a few specific functionalities needed for visual crowd research. 

A p5.js sketch is a JavaScript file that uses the commands and structure of the  Processing language. In the JavaScript file, images and data are loaded, screen presentation defined and data is collected from mouse movement and keyboard.  p5.js even makes it possible to use smartphone events (like orientation and acceleration) as input with a single function call.

Hosting online experiments and collecting data can be done in three ways: via AMT, via a server, or locally. The latter is not entirely `online', but in some cases it can be useful to send a link to a p5.js sketch and save data locally on the participants' computer, for example for student projects. This is relatively trivial and will not be discussed further. Of the other two options we first discuss the AMT route, as we did all our experiments in that way. 

When using the AMT requester portal, the experimenter can choose a template to start coding their own experiment. The HMTL, CSS and JavaScript code for a project can be written or copied into the editor. AMT has created a library of custom HTML tags (Crowd HTML Elements) that can be used to save data. The fact that saving data on AMT makes use of HTML elements while p5.js makes use of JavaScript data structure is the first small challenge to be overcome. We found that collecting data in a p5.js Table object is most convenient. When the experiment finishes, the Table contents are then transferred to the HTML tag using a simple function. 

Besides saving data, the experimental design data needs to be loaded. Often a data file is needed that contains sampling points, image names, etc. Using a CSV file and importing it as a Table is most convenient. The order of the trials can be randomised to counterbalance order effects. When linking images and scripts from other websites, problems with `Cross-Origin Request Sharing' can occur. We found that hosting on Amazon S3 worked well and the relevant settings could easily be adjusted. 

An advantage of conducting everything through AMT is that hosting and participant recruitment is integrated. This can also be a disadvantage if participants are recruited outside AMT. In that case, a web-server is needed to save the experimental data. As the maintenance of servers is relatively complex, we would shortly like to share an alternative of so-called `serverless' hosting. AWS Lambda is a platform of Amazon Web Services (AWS) that allows for serverless solutions. By creating a Lambda function it is possible to write data directly from a p5.js sketch to AWS S3 storage. In other words, when a server is needed, AWS Lambda can be used instead to drastically reduce time and effort, as well as reducing the skill threshold. In the appendix we will share a short instruction, and in the repository we share more elaborate information.

\subsection{Experiment 1: Change blindness}

\subsubsection{Participants}
A total of 30 participants were recruited on AMT. The data of 3 participants was left out as their average reaction time was 3 seconds (which is suspiciously short for this paradigm) and clicks appeared at random locations. 
\subsubsection{Procedure}
\begin{figure}[!h]
\begin{center}
\putfigman{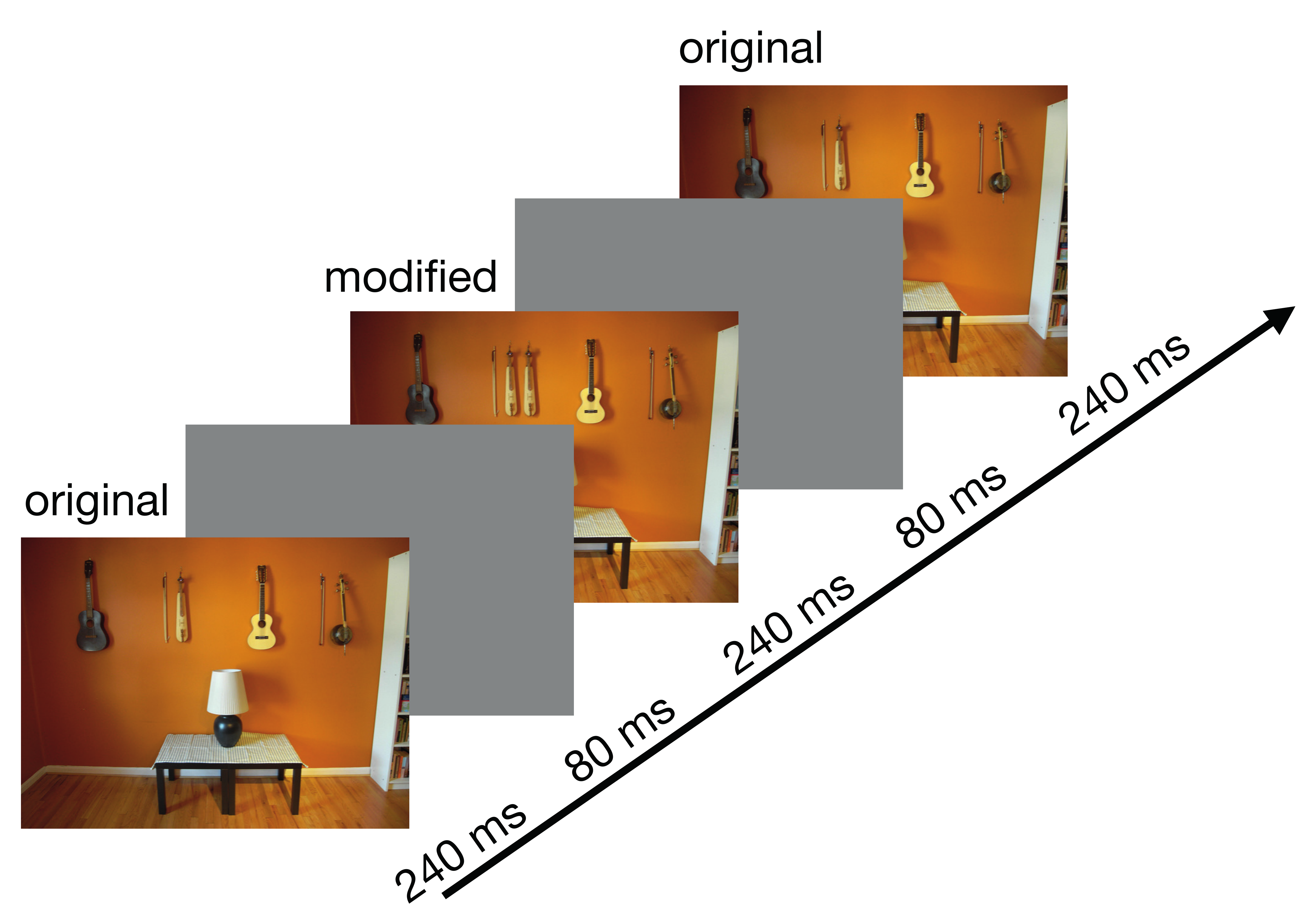}{.9}
\caption{Overview of the changeblindness paradigm, using stimuli from \cite{Ma13Tvcg}. The original and modified version are shown in succession with an intermediate gray frame. }
\label{changeblindnessPicture}
\end{center}
\end{figure}

The experimental procedure is visualised in figure \ref{changeblindnessPicture}: observers saw a flickering image that alternated between two versions and intermitted by a grey screen. In one image, an object disappeared or moved relative to the other image. Observers had to click as fast as possible on the location of that object. 
\subsubsection{Images}
For our change blindness experiment we used stimuli from \cite{Ma13Tvcg}. We used 10 picture pairs from their original set, 5 'easy' picture pairs with average reaction times of about 5 seconds and 5 'difficult' pairs with average reaction times of about 60 seconds. These different reaction times make them interesting and challenging to put on AMT: would a participant be willing to spend 60 seconds to find a difficult target?

\subsubsection{p5.js}
Two aspects of the p5.js sketch for the change blindness paradigm are interesting to discuss. First, the loading of images is done asynchronously, meaning that the code will continue to execute, before the data or images have finished loading. While this is useful for normal web pages, in the case of web experiments, this can lead to serious issues. To prevent these problems, it is possible to use callback functions in p5.js, which will only execute once the data has finished loading. This is typically needed for experiments with a set of images. Secondly, we used the frameRate() function to control the flicker paradigm. In the original publication \cite{Rensink1996} an image was shown for 240 ms and the gray blank for 80 ms. Thus, using a frame rate of 12.5 Hz and assigning 3 frames to the image and 1 frame to the blank will result in the required visual presentation. The logging of mouse clicks is not affected by this frame rate setting.

\subsection{Experiment 2: Bubble View}

\subsubsection{Participants}
Per image, 9 observers participated in the experiment. This is perhaps on the low side, but knowing that the click patterns are rather robust \cite{Kim2017a}, we thought this was sufficient. We used the following criteria for the workers: located in the US, acceptance rate of 95\% or higher for previous HITs and more than 1000 accepted HITs. 
\subsubsection{Procedure}
\begin{figure}[!h]
\begin{center}
\putfigman{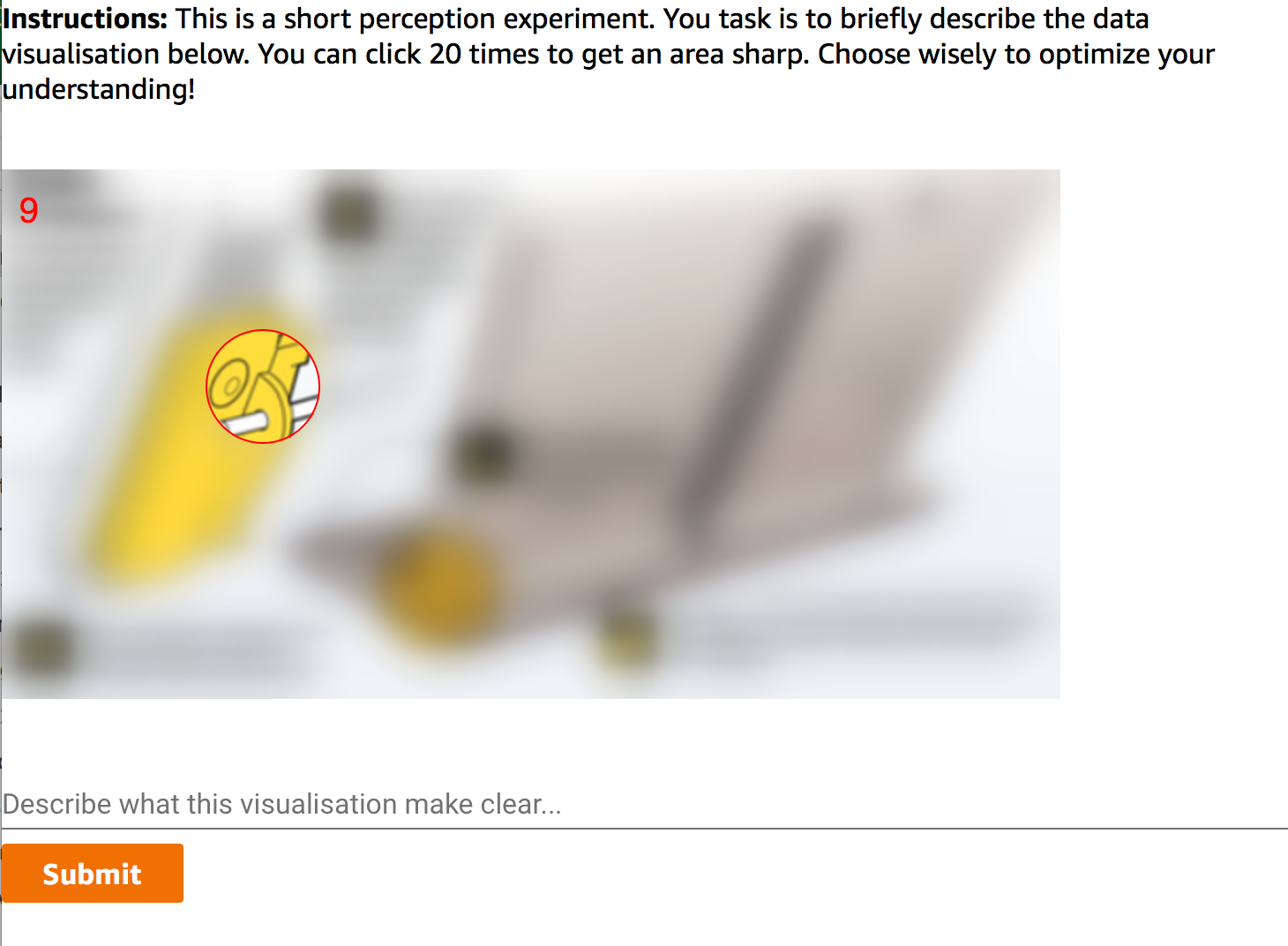}{1}
\caption{The BubbleView interface in AMT. The instruction text, text input and submit button are all in the AMT HTML, while the image interaction is in p5.js.}
\label{bubblemethod}
\end{center}
\end{figure}

A screenshot of the experiment is shown in figure \ref{bubblemethod}. Participants had to describe what the blurred image conveyed and were allowed to click 20 times to sharpen those areas. The `bubble' had a radius of 32 pixels; the image width was held constant at 600 pixels.

\subsubsection{Images}
We used two images of the data visualisation set that were used in the original BubbleView paper \cite{Kim2017a}.

\subsubsection{p5.js}
The p5.js library allows for basic image filter operations such as blur, and gives access to pixel values through the pixels() command. In our implementation of BubbleView, we showed a blurred version of an image and defined a sharp aperture by displaying the pixels from the (sharp) original image around a mouse click location. These are all relatively trivial steps in p5.js. What is furthermore interesting in this example is that we mixed a p5.js sketch with an crowd HTML element for the textual input. As AMT's crowd HTML elements are a very suitable solution for collecting textual input, we used that instead of p5.

\subsection{Experiment 3: 3D shape perception}

\subsubsection{Participants}
A total of 20 workers participated in this study. As selection criteria, we used so-called 'masters', according to AMT ``specialized group of Workers who consistently demonstrate accuracy in performing a wide range of HITs''. 
\subsubsection{Procedure}
\begin{figure}[!h]
\begin{center}
\putfigman{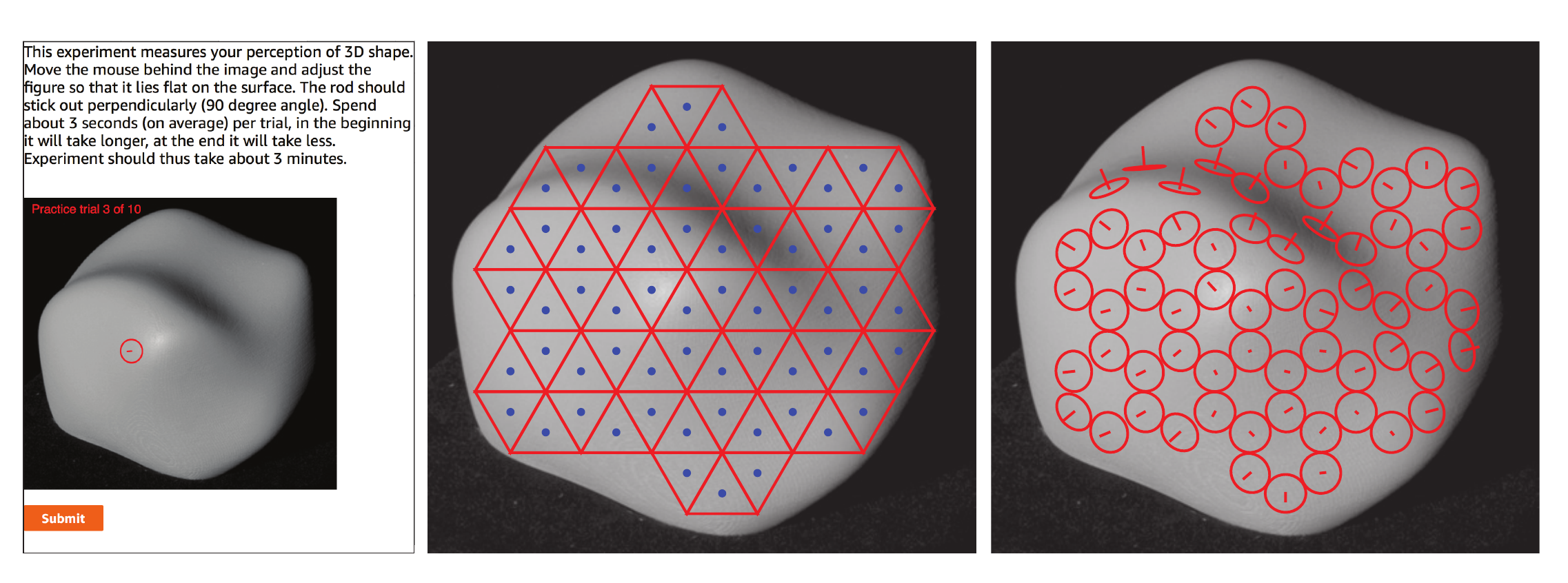}{1}
\caption{Left the experiment screen is shown with the explanation. In the middle the triangulation can be seen, which is used for reconstructing the global depth from the local settings. The blue dots are the barycentres of the triangles and are the sampling points where the gauge figure appeared (in random order). On the right, all settings of 1 observer are shown. Mind that the observers did not see more then one setting simultaneously.}
\label{gaugefiguremethod}
\end{center}
\end{figure}

The experimental task is to adjust the 3D orientation of a figure that looks like a drawing pin or thumb tack. An example trial is shown in figure \ref{gaugefiguremethod}. Observers had the adjust the orientation such that the figure appeared to lie flat on the pictorial surface, with the stick pointing perpendicularly outwards. Participants were instructed to spend no more than 3 seconds per trial. 

\subsubsection{Image}  
We used a stimulus from \cite{Wijntjes2012a} but with a courser sampling of 64 points, which is approximately recommended by a previous AMT study \cite{cole}
\subsubsection{Data analysis}
There are various ways to analyse the data, for example, simply correlate slant (out-of-plane rotation) or tilt (in-plane rotation) settings among observers. A more complex way is to reconstruct the global 3D surface by integrating the settings because these are essentially the local attitude (the derivative) of the surface. We will not discuss these details further but choose the reconstruction method primarily because it nicely visualises the results. We used \cite{wijntjes2012probing} to setup the triangulation and reconstruct the 3D relief.

\subsection{Experiment 4: Composition preference}

\subsubsection{Participants}
A total of 100 participants were recruited at AMT. 
\subsubsection{Procedure}
\begin{figure}[!h]
\begin{center}
\putfigman{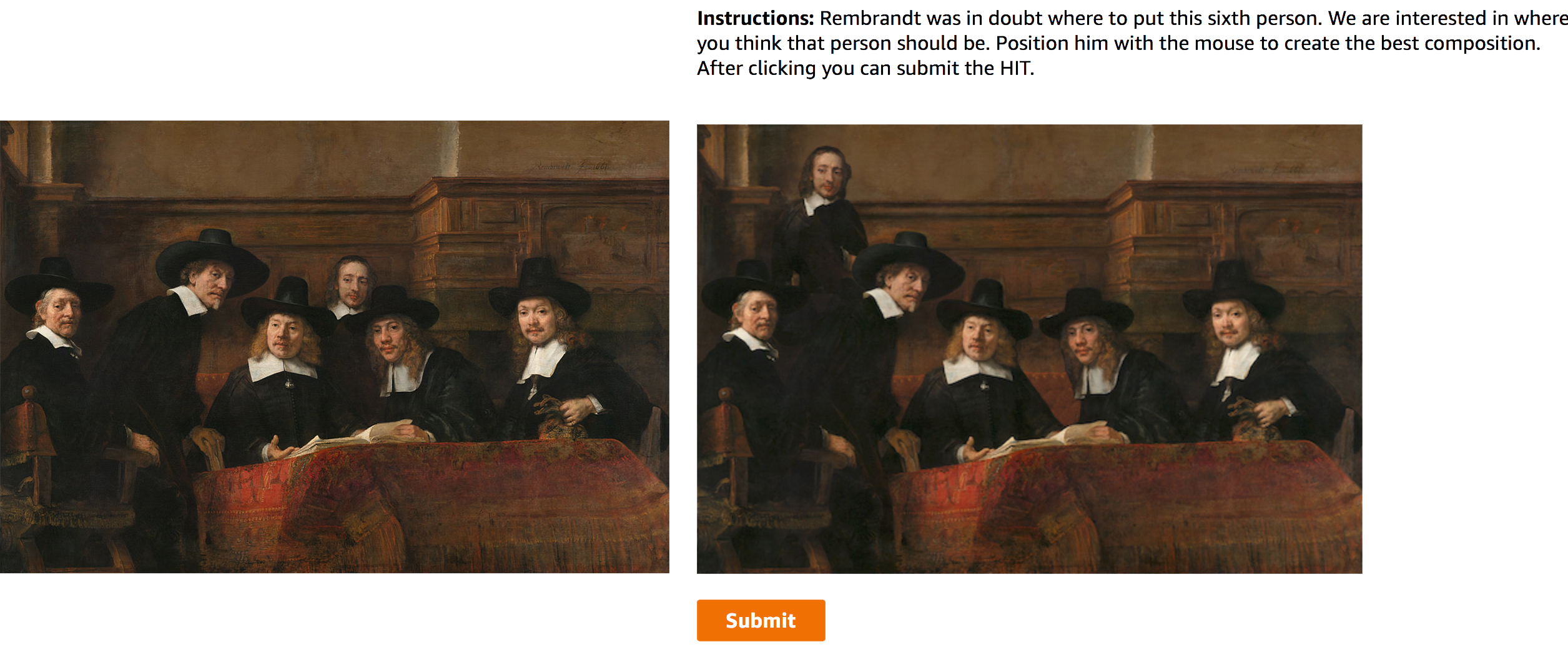}{1}
\caption{Left the original painting is shown: the `Syndics of the Drapers Guild' (1662) by Rembrandt. On the right, the experimental interface is shown. As can be see: the servant is cut out of the original and can be freely positioned anywhere in the canvas. Painting reproduced from The Rijksmuseum, The Netherlands under a CC 0.0 licence.  }
\label{compositionmethod}
\end{center}
\end{figure}

Participants were instructed to position the servant at the location that they believed resulted in the best composition. The experiment is shown in figure \ref{compositionmethod}. The initial position of the servant depended on where the users' mouse was at that moment, and was thus not actively randomised.

\subsubsection{Image} On the left side of Figure \ref{compositionmethod} the original (unedited) image of `Syndics of the Drapers Guild' is shown. We used photo editing software to cut out the foreground scene and independently the servant. The place of the servant was filled with image elements of the remaining scene as to not give away Rembrandts' choice. 
\subsubsection{p5.js}
The standard function image() was used to display the foreground, background and make the servants' position depend on the current mouse position.

\subsection{Experiment 5: Perspective elevation}

\subsubsection{Participants} The experiment was split in three blocks for which each 10 participants were recruited. 
\subsubsection{Procedure}
Participants were instructed to first annotate where they saw the horizon by adjusting a red horizontal line. Then they had to draw lines between the feet and head for about 10 to 15 people in the scene. They were encouraged to choose people that were at various distances, including those far away (and tiny on the screen). 

\subsubsection{Data analysis}
A linear regression of the vertical distance between the feet and the horizon as a function of human sizes was performed. The slope of the model has a direct meaning: it is the height of the perspective viewpoint in terms of human sizes. For example, if the slope is 2, the viewpoint of the painter was (virtually) 2 human lengths (approximately 3.5 meters) tall. We set the offset to zero for the actual elevation data, but also used it as an free parameter and found little difference (blue and orange lines in figure \ref{avercamp}). 

\subsubsection{Images} We first downloaded all Avercamp paintings we could find from WikiArt. Then we selected those that showed ice and multiple people. We used the WikiArt meta-data for production years.

\section{Results and discussions}

\subsection{Change blindness}
\begin{figure}[!h]
\begin{center}
\putfigman{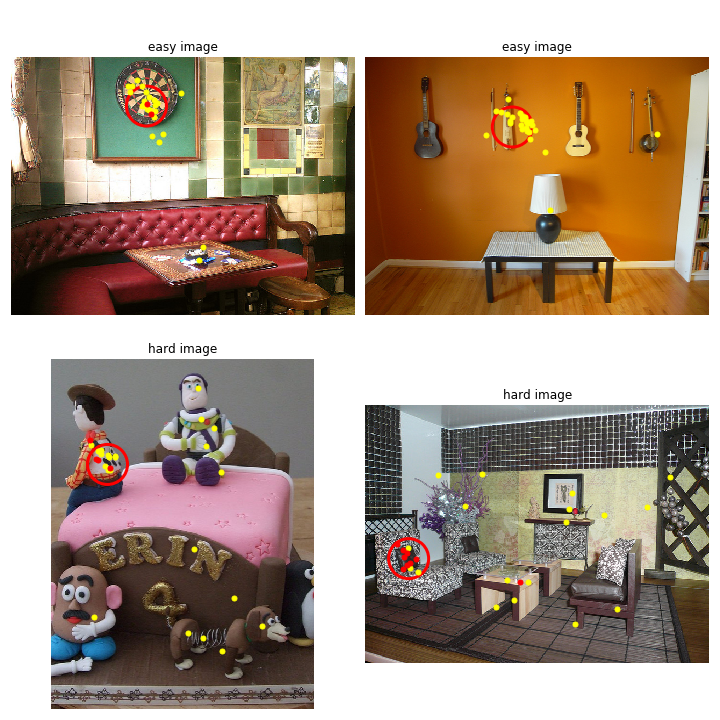}{.8}
\caption{4 examples of images used for the change blindness experiment. The top images were designated easy and the bottom as hard, as per \cite{Ma13Tvcg}. The red ellipse indicates the change target location. Yellow and red clicks were made by participants. Red clicks were placed after one minute, when the target location was revealed to participants }
\label{changeBlindImages}
\end{center}
\end{figure}

Four of the 10 images used in the change blindness experiment are visualized in figure \ref{changeBlindImages} with the location of the target, as well as the location the participants clicked. The clicks are visualized as the yellow points in \ref{changeBlindImages}. To save participants from frustration, if they did not make a click within one minute, we would highlight the target. Clicks made after this one minute are visualized in red. As can be seen in figure \ref{changeBlindImages}, the majority of clicks placed by participants are at, or near the target, with a smaller number of clicks located far from the target. These distant clicks could be the participants wrongfully believing they saw a change, or the participant giving up. 

We looked at the reaction times for each trials that qualified the following two conditions: 1) clicks were made within one minute (to remove trials in which the correct answer was shown to the participant) and 2) if their click was placed correctly. \textit{Correctly} here is defined as being within a 0.1 radius from the changing object, based on an image re-scaled to a width of 1 by 1.  Note that this is a rather conservative criterion, as in some trials a relatively large object moves around in which where clicks made at the center of the changing object would qualify, but clicks at the edges might not.  Performing an independent t-test on the reaction times confirmed that the easy images (\emph{M} = 10.7s, \emph{SD} = 14.1s) were found significantly faster than the hard images (\emph{M} = 25.1s, \emph{SD} = 23.6s), \emph{t}(269) = -6.08, \emph{p} $<$ .0001. \cite{Ma13Tvcg} found average reaction times of 5.1 seconds for easy images and 58.7 seconds for the difficult images. Thus, our difference is substantially smaller. This is  surprising, but there can be various reasons for this difference, for example images size: we used rather small images.  \cite{Ma13Tvcg} did not report image size but it was likely larger. Furthermore, we clipped the experiment at 60 seconds, to not frustrate the participants. This obviously affects the mean reaction time to be substantially lower.

\subsection{Bubble View}
\begin{figure}[!h]
\begin{center}
\putfigman{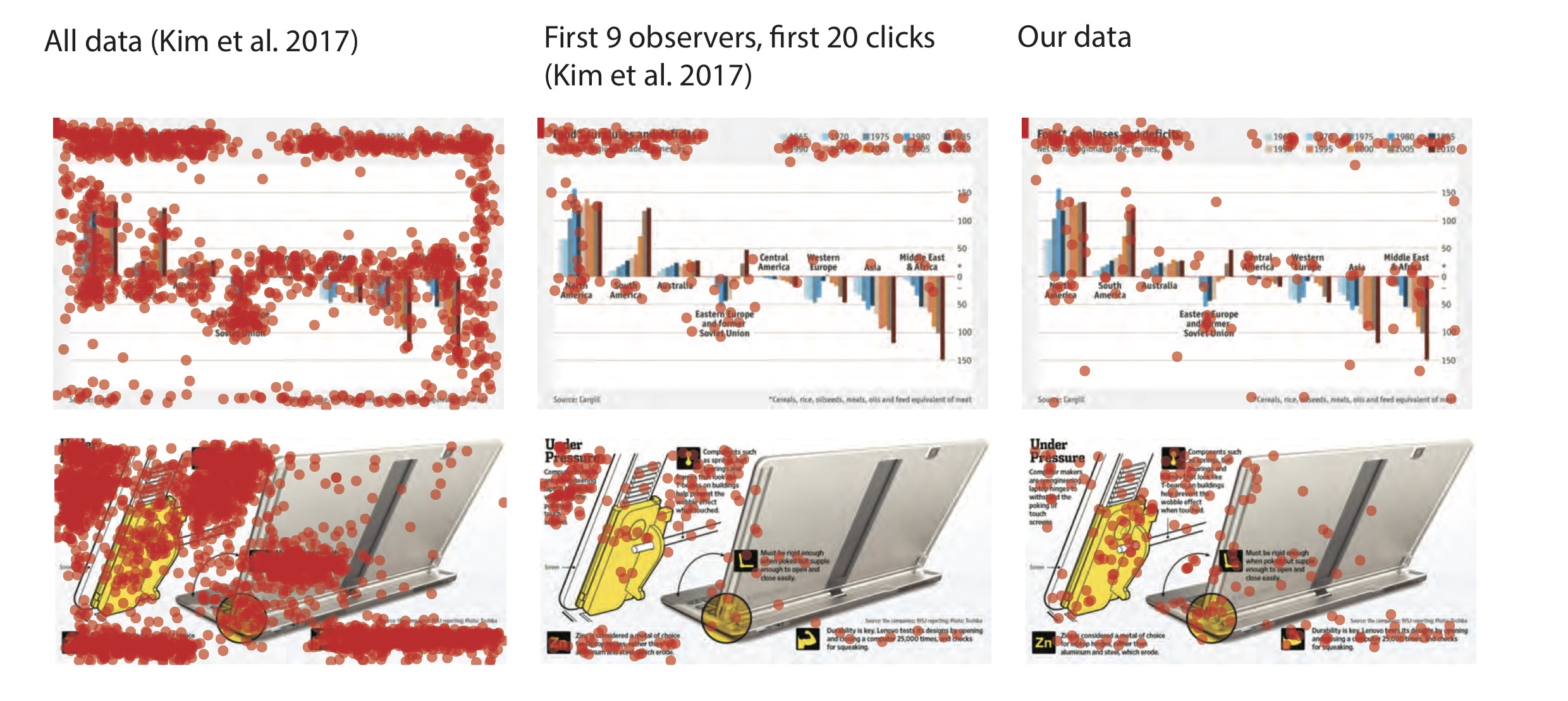}{1}
\caption{Two experiments compared with the original data from \cite{Kim2017a}. Because we only used 9 participants and let them click 20 times we imposed the same restriction on the original data by selecting their first 9 observers and 20 clicks.}
\label{bubble}
\end{center}
\end{figure}

We plotted the raw data in figure \ref{bubble}. As the data from \cite{Kim2017a} differed from ours by the number of participants and clicks, we corrected for that by visualizing only their first 9 observers and 20 clicks per observer in the middle column. It should be noted that this filtering may bias the results because click strategies may depend on either time or click limitations. 

The pattern of clicks seems rather similar: participants mostly click on text. In the bottom row the click data also seems similar except that the yellow element attracts more attention in our experiment than the original. It is a zoomed in picture of the screen flipping mechanism which is relatively difficult to visually understand (even if you see the sharp version). The yellow part may need multiple clicks to understand its global appearance, while the text between the yellow part and the laptop is simply too much to reveal by clicking.

\subsection{Gauge figure attitude probe}

\begin{figure}[!h]
\begin{center}
\putfigman{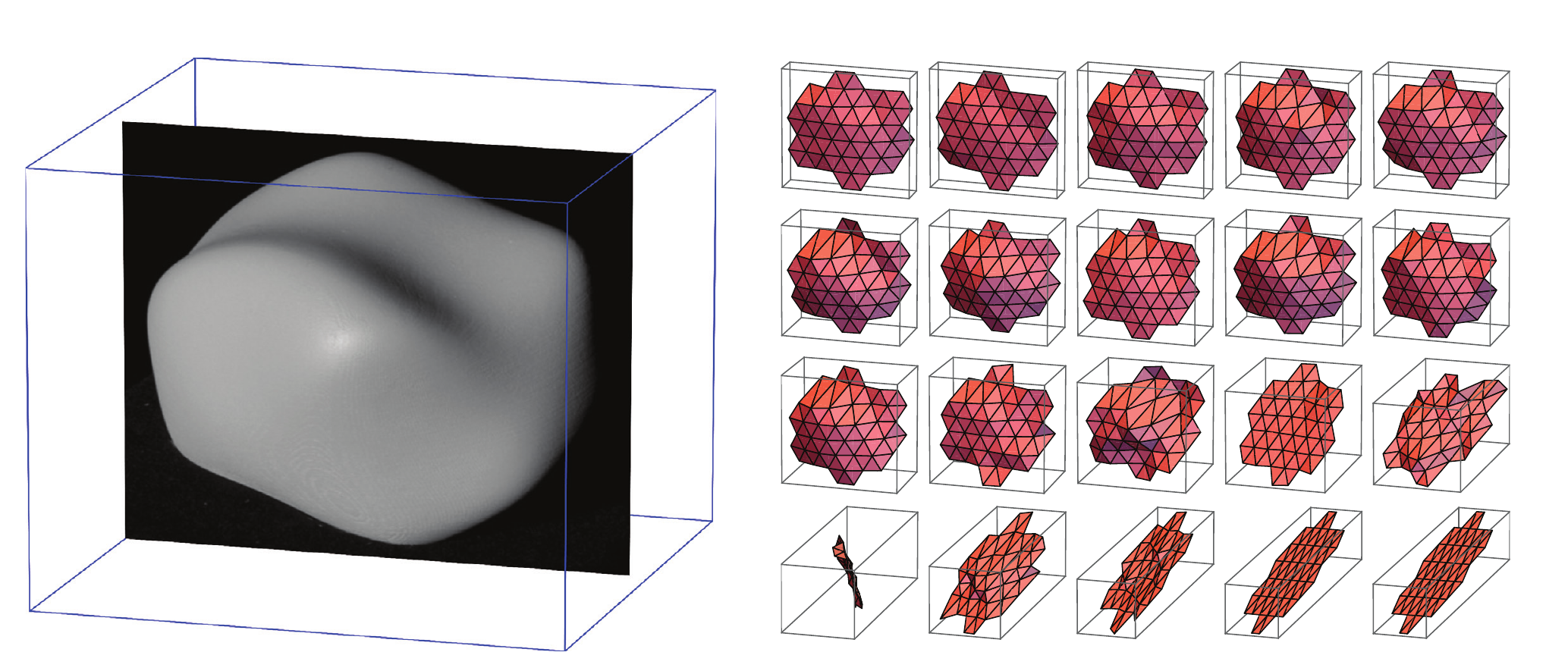}{1}
\caption{On the left, the stimulus is shown, aligned in a 3D frame while the subjective surface reliefs are visualised (on the right), which are ordered from shallow to deep (attend to the boxes to see this clearly). As can be seen in these results, there is quite some variability in the perception for 3D shape, and also in understanding the task (e.g., the last 7 observers).  }
\label{gaugefigureresults3}
\end{center}
\end{figure}

We reconstructed the 3D surfaces on the basis of local surface attitude estimates of 20 participants. The data is shown in figure \ref{gaugefigureresults3} and ordered by relief depth. What can immediately be seen is that about 35\% (the last 7) of the observers did not seem to understand the task. Although there is not much known in the literature about this, we know from experience that the instruction of gauge figure experiments in the lab requires substantial attention. It  seems relatively difficult to understand although showing visual examples generally helps. We have also experimented with instruction videos in other (for now unpublished) AMT research, which seemed to improve understanding. Taking into account that in this case we only used 64 words to describe the complete task puts the results in perspective: having 65\% 'normal' data is actually above our own expectation. 

Although there is further analysis possible, e.g., quantify how integrable (globally consistent) the attitude estimations are \cite{koen1992} or whether differences can be described by affine transformations \cite{Koe2001}. However, we think it can also be visually inferred that the depth range and global attitude seem to vary substantially between observers which is much in line with previous findings. From an annotation point of view, these results imply caution. It confirms the often found individual differences in \emph{perception} \cite{Koe2001} affect the reliability of   depth annotations for machine learning such as \cite{Chen_2020_CVPR}.

\subsection{Composition}

\begin{figure}[!h]
\begin{center}
\putfigman{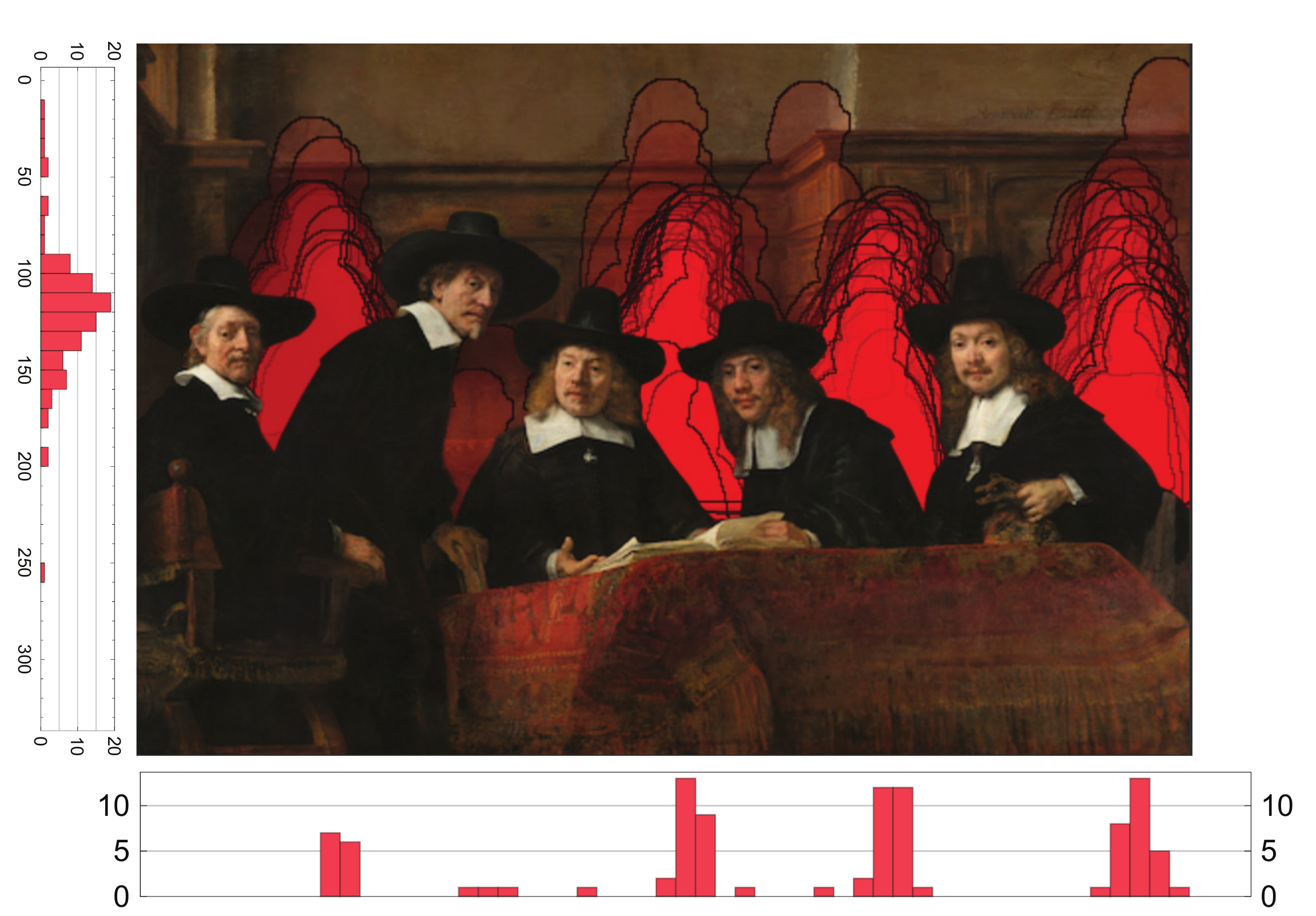}{.7}
\caption{Visualisation of 100 responses of where the servant would result in the best composition, according to participants. There are 4 dominant horizontal positions with more or less equal probability, except the one on the left.}
\label{staalmeesterresults}
\end{center}
\end{figure}

The compositional preferences of 100 participants are visualised in figure \ref{staalmeesterresults}. It can be readily seen that 4 horizontal locations dominate the data. One of them is similar to the actual painting, and the far right alternative is similar to Rembrandts' under drawing, i.e., the initial `sketch' \cite{VanSchendel1956}. Perhaps the other two locations have also been considered by Rembrandt. 

A natural follow-up question for this short experiment is clearly what the experiential (i.e., perceptual, aesthetic and/or emotional) effect these choices would have.

\subsection{Perspective}
\begin{figure*}[!t]
\centering
\subfloat[]{\includegraphics[width=\columnwidth]{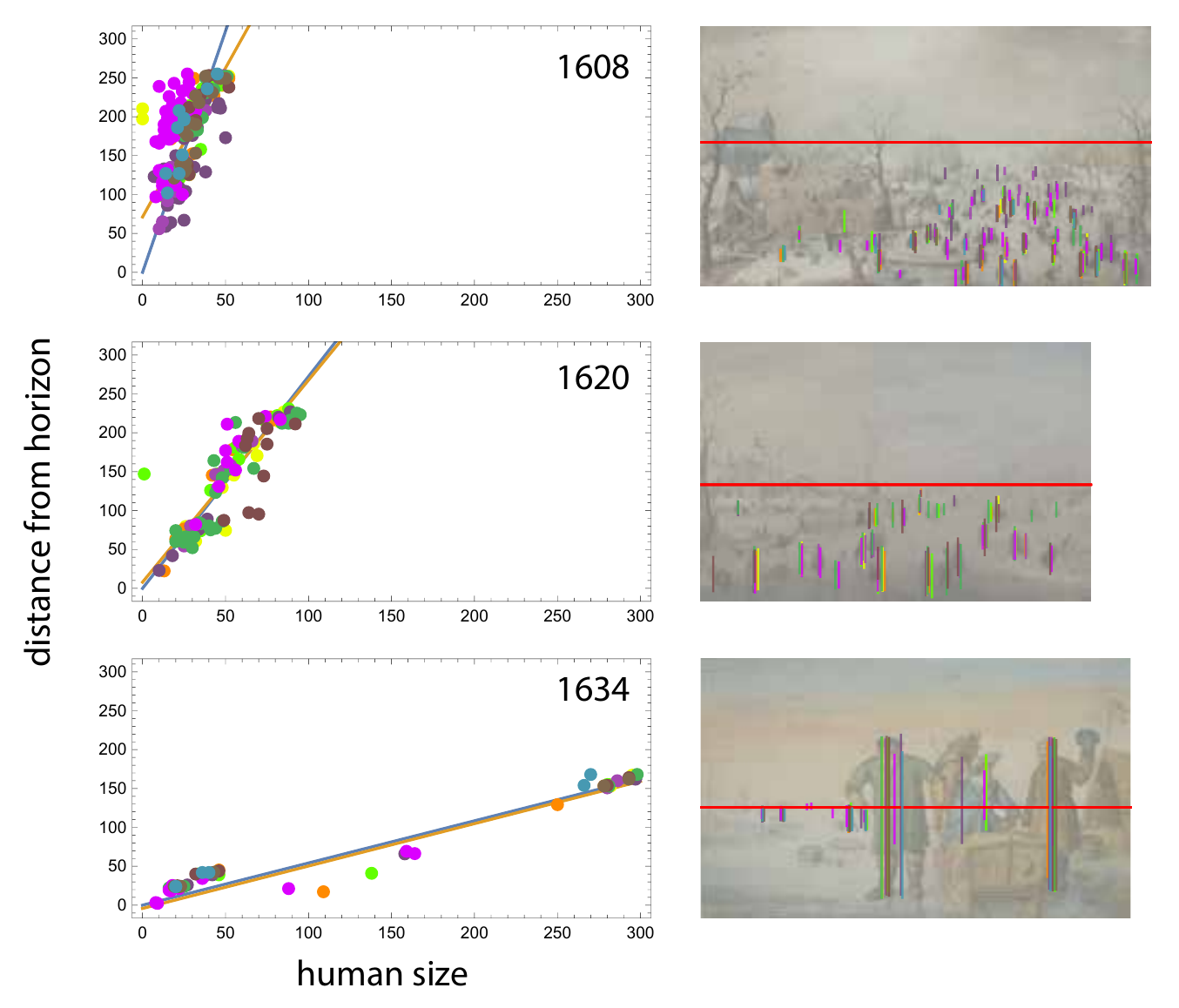}%
\label{fig_first_case}}
\hfil
\subfloat[]{\includegraphics[width=\columnwidth]{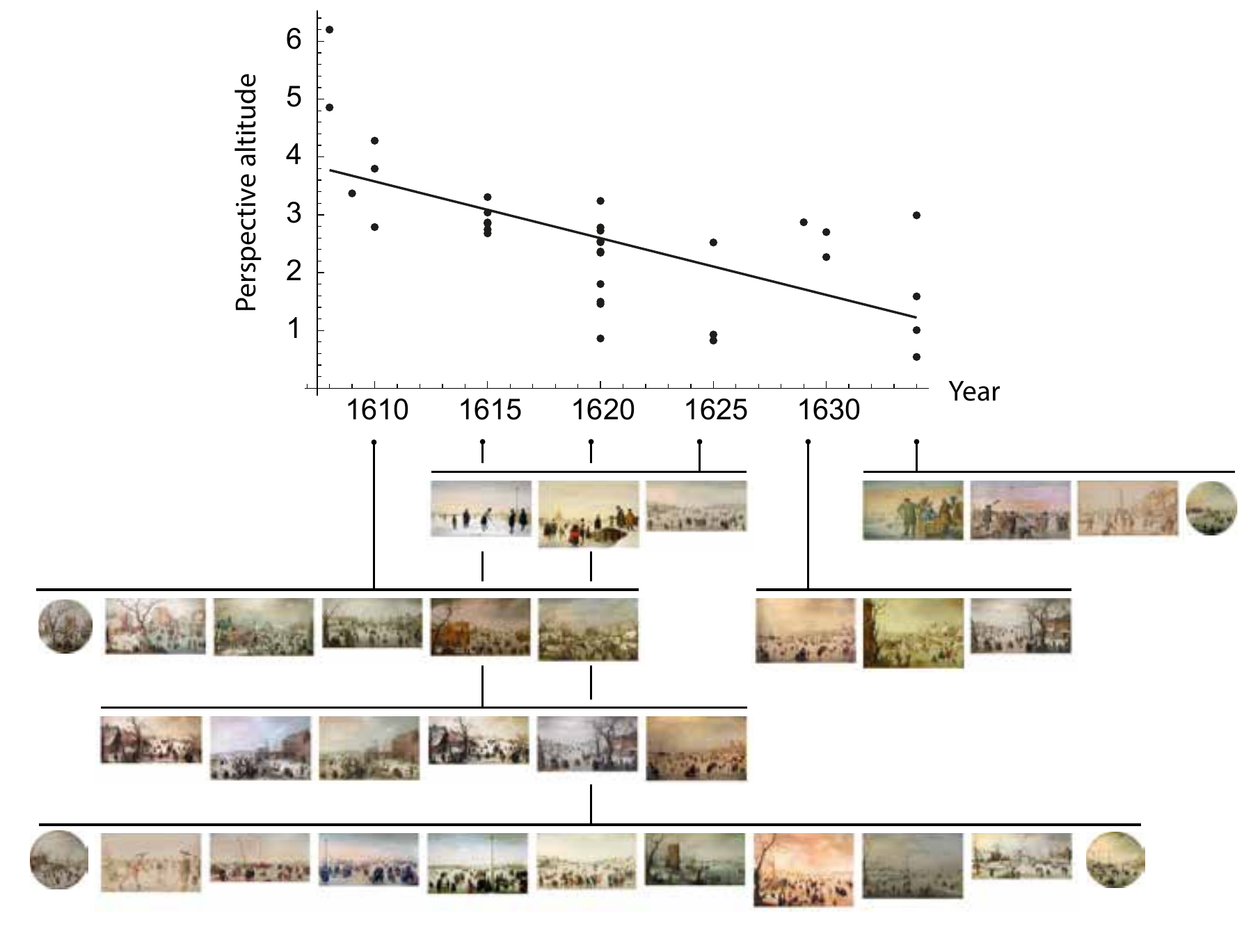}%
\label{fig_second_case}}
\caption{(a) Three examples of the data. The (reduced contrast) paintings are overlaid with the raw annotation data. Different colours mean different annotators. The slope in the scatterplots indicate elevation of the viewpoint. (b) Here, the viewpoint elevations are plotted over time. All images are shown under the years. As can be seen, a negative trend is clearly visible.}
\label{avercamp}
\end{figure*}

In figure \ref{avercamp}, 3 representative examples are displayed on the left, and the overall perspective elevation over time is shown on the right. Before discussing these results, it should be reported that horizon estimation turned out to be rather fluctuating and seemed to arise from much confusion. Apparently, the concept of horizon is very clear to the authors, but not to the general AMT worker, or at least not in the terms that we used to explained it. For further analysis, we choose one representative participant of which we used the horizon data for further computations.

The negative slope (-0.10) was highly significant ($t(32)=-4.84405$, $p<0.0001$ and implies that Avercamp used a lower perspective elevation as he became older. The effect is rather strong, and we did not anticipate this effect. Interestingly, the same trend is found for Canaletto \cite{Wijntjes2018AnnotatingSH}, an Italian painter who was active a century after Avercamp. As these two painters seem unrelated, it raises the question whether there is a general trend among artists when it comes to perspective elevation and age.

\section{General discussion}

\subsection{p5.js}
We have demonstrated picture-centered research using p5.js for visual crowdsourcing. Many functionalities that are needed for visual crowdsourcing are available in p5.js, which is due to the shared interest of artists, designers and vision researchers in both accuracy and versatility. We replicated three experimental paradigms and demonstrated two new experiments. That the results are similar to previous findings may not be too surprising, although two out of three experiments were originally conducted offline. The fact that we find rather similar results adds to the evidence that crowdsourced experiments can replicate lab experiments. 

More interesting than the experimental results per se is the fact that we demonstrated a simple and effective way to perform a wide variety of visual crowdsourcing experiments. With the BubbleView experiment, we demonstrated the ability to filter (blur) images and display specific pixel regions (sharp areas) and how to collect mouse click data. The Change Blindness experiment exemplified how to make use of external measurement lists and how to measure reaction times. The 3D shape experiment showed how to use shape primitives and how to load and use sample locations. The Composition experiment showed the usefulness of images with transparent areas. Overall, this set of  experiments serves as an adequate basis for developing new experimental paradigms. 

While the use of p5.js can also stand on its' own, we specifically investigated it in the context of Amazon Mechanical Turk. While collecting information about other AMT studies, we noticed that although many researchers share their code publicly, it cannot always be easily reused. For example, BubbleView \cite{Kim2017a} released their code but not the integration with AMT because that requires \emph{``more complex development settings including a database, a web server, and scripts for automatically managing AMT HITs.''}. These are perfectly valid reasons for not offering a simple way to replicate the experiment on AMT. But it does increase the threshold for less experienced researchers to use their experimental paradigm.

Similar to the efforts of both Processing and p5.js in making programming accessible to a large audience, our aim was to increase accessibility of visual crowdsourcing research. The solution we studied works well but also has its limitations. For example observer management and complex experimental designs may be easier in psiTurk \cite{Gureckis2016}, PyschoPy \cite{Peirce2019} or TurkPrime \cite{Litman2017}, while fast and accurate graphics may work better with custom code and servers. Moreover, database management, adaptive and automatic HIT creation require more complex solutions. Yet, these trade-offs come with the advantage that designing and running an experiment is as simple as copy-pasting a p5.js sketch from the online editor to the AMT requester portal. 

\subsection{Picture-centered research}

There is a reciprocity in the existing experimental paradigms. Experiments involving pictures and observers reveal something about the relationship, but also about the pictures and observers themselves. This reciprocity is rarely used. For example, material annotations in \cite{bell2013opensurfaces} offer potentially rich behavioural data as annotators  quantified surface reflectance similar to perception studies (e.g., \cite{pellacini2000toward}). Vice versa, experiments concerning attention and saliency can also be used to investigate an individual or a series of pictures. In design, this is already practiced. Change Blindness and BubbleView experiments are used in visual information design \cite{Varakin2004,Newman2020}. Furthermore, annotations are starting to get used in areas where large collections of cultural data are studied, such as digital art  \cite{Wijntjes2018AnnotatingSH}.

Pictures are means of communication. The study of sender, receiver, medium, code, context, etc all belong to picture-centered research. Some of these topics can be studied quantitatively, making use of those for whom pictures are intended: the human receiver. We aimed to increases the accessibility of  visual crowdsourcing and demonstrate its usage, both technically and conceptually, and look forward to all unforeseen future investigations that may come from this.

\section*{Acknowledgements}
This work is part of the research program Visual Communication of Material Properties with project number 276-54-001, which is financed by the Dutch Research Council (NWO). Furthermore, we would like to thank students from the course Visual Communication Design at the TU Delft, who enthusiastically joined our fascination of visual perception and communication and caused a type of reciprocal inspiration. Lastly, we would like to thank prof. Joris Dik, who introduced us to many intriguing art historical questions of which the Drapers Guild is merely one. 

\section*{Appendix}
Here, we report code fragments that are important for running p5.js crowd experiments. If pictures are used they need to be hosted at a server that has the correct CORS settings. Image loading needs to be done in the preload function, or at least it needs to be guaranteed that images are loaded before initial presentation. Here is an example where one image is loaded. For an example where multiple images are loaded on the basis of the filenames in a .csv file, please see the GitHub repository https://github.com/maartenwijntjes/p5-sketch-and-test. 

\begin{lstlisting}
let path = 'https://materialcom.s3.eu-central-1.amazonaws.com/bubbleview/stims/'
let imageNameShort = 'wsj104.png';
let imageName = path + imageNameShort;
function preload() {
    image = loadImage(imageName);
}
\end{lstlisting}

In the online editor, the p5.js canvas is automatically attached to an HTML element. Outside of the editor, this has to be done manually, therefore it is recommended to use the following with the setup() area:

\begin{lstlisting}
canvas = createCanvas(sketchWidth, sketchHeight);
if (!onP5Editor()) {
    canvas.parent('p5sketch');
}
 \end{lstlisting}
where 
\begin{lstlisting}
function onP5Editor() {
    return document
            .location
            .ancestorOrigins[0]
            .includes('editor.p5.js.org')
}
\end{lstlisting}

Now, the `p5sketch' tag can be used in the HTML code, for example like 
\begin{lstlisting}
<div id="p5sketch"></div>
\end{lstlisting}

For data collection it is convenient to use a p5.js table with a manually defined header:
\begin{lstlisting}
let header = ['x', 'y', 'r', 'imageName'];
data = new p5.Table();
for (let i = 0; i < header.length; i++) {
    data.addColumn(header[i]);
}
\end{lstlisting}

To save the collected data, either locally or via AMT, we used:

\begin{lstlisting}
function finished() {
    if (onP5Editor()) {
        saveTable(outputData, 'data.csv');
    } else {
        experimentOutput = document.getElementById('experimentOutput');
        experimentOutput.value = table2csv();
  }
}
\end{lstlisting}

Which assumes the following HTML element exists:

\begin{lstlisting}
<crowd-input hidden name="dataOutput" id="experimentOutput" required>
</crowd-input>
\end{lstlisting}

The finished function is automatically called when the experiment is finished (e.g., when the number of trials or total time is reached). As can be seen, we use the P5Editor function defined previously. If the sketch runs in the online editor, the p5.js function saveTable is used to convert the p5.js Table object to a .csv file. If the sketch runs on AMT, the p5.js Table is converted to csv text using our custom table2csv function. The csv text is then inserted into an AMT crowd-input HTML element, which saves the data on the AMT platform.

Lastly, we share a function for sending data to an S3 bucket in a `serverless' fashion, which can be practical for remote experiments outside AMT. Please first visit https://github.com/aws-samples/s3-to-lambda-patterns and follow the instructions from the ``Serverless File Uploads using S3 and Lambda'' demonstration. The function below refers to an API link that refers to the Lambda function that will create a temporary link with the correct permissions to upload (PUT) your data (testdata).

\begin{lstlisting}
function transferdata2bucket(){
  var request = $.ajax({
    method  : "GET",
    url : 'https://YOURKEYHERE.execute-api.eu-central-1.amazonaws.com/default/getPresignedURL'
  })
  
  request.done(function(response){
    uploadURL=JSON.parse(request.responseText)['uploadURL']
    request2 = $.ajax({
      url : uploadURL,
      data : testdata,
      type : "PUT",
      contentType : "text"
    })
    request2.done(function(){
       print('Success! Data was send.')
     })
  })
}
\end{lstlisting}

\bibliographystyle{JPercepImag}
\bibliography{sketch-and-test}
\end{document}